\documentclass[showpacs,preprintnumbers,amsmath,amssymb]{revtex4}

\usepackage{graphicx}
\usepackage{bm}

\begin{document}

\title{Existence and non-existence of breather solutions  in damped and driven nonlinear lattices}

\author{D. Hennig}
\affiliation{Department of Mathematics, University of Portsmouth, Portsmouth, PO1 3HF, UK}

\date{\today}

\begin{abstract}
\noindent 
We investigate the existence of spatially localised 
solutions, in the form of discrete breathers, in general damped and driven nonlinear lattice
systems of coupled oscillators. Conditions for the exponential decay of the 
difference between the maximal and minimal amplitudes of the oscillators are provided which  proves that 
  initial non-uniform  spatial patterns representing breathers  
attain exponentially fast a spatially uniform state preventing  
the formation and/or preservation of any breather  solution at all. 
Strikingly our results are generic in the sense that they hold for arbitrary 
dimension of the system,  any attractive interaction,
coupling strength and on-site potential and general driving fields. 
Furthermore, our rigorous quantitative results establish 
 conditions under which discrete breathers in general damped and driven
nonlinear lattices can exist at all and 
open the way for further research on the emergent dynamical scenarios, in particular features of pattern formation, localisation and 
synchronisation, in coupled cell networks.  
\end{abstract}

\maketitle

\noindent Intrinsic localised modes (ILMs) or discrete breathers in nonlinear lattices
have  attracted significant interest recently, not least due to the important role they play in many 
physical realms where features 
of localisation in systems of coupled oscillators are involved (for a review see \cite{Flach1} and references therein),\cite{Aubry}-\cite{PhysicsReports}. 
For conservative systems proofs of existence and (exponential) stability 
of breathers, as spatially localised and time-periodically varying solutions,   were 
provided in \cite{MacKay} and \cite{Bambusi} respectively. Analytical and numerical methods have been
developed to continue breather solutions in conservative and dissipative systems  
starting from the anti-integrable limit \cite{Sepulchere},\cite{Marin}. During recent years the
existence of breathers has been verified in a number 
of experiments in various contexts including micro-mechanical cantilever arrays \cite{cantilever}, 
arrays of coupled Josephson junctions \cite{josephson}, coupled optical wave guides \cite{optical}, 
Bose-Einstein condensates in optical lattices \cite{BEC}, in coupled torsion pendula \cite{Jesus}, 
electrical transmission lines \cite{electrical}, and granular crystals \cite{crystals}. 
Regarding their creation mechanism in conservative systems, modulational instability (MI) provides the route to 
the formation of breathers originating from an initially spatially homogeneous state imposed to (weak) perturbations. 
To be precise, the MI of band edge plane waves 
triggers an inherent instability leading to the formation of a spatially  localised state 
\cite{Remoissenet}. Departing from this often too ideal assumption of a conservative system 
requires, for more realistic models, the inclusion of dissipation.  Accomplishing breather solutions in the 
presence of dissipation requests some 
compensating energy injection mechanism. As far as  their persistence is concerned 
it is expected that breathers can be continued from a conservative system  into a system augmented by  
weak dissipation  and driving. Compared to their Hamiltonian (conservative) counterparts breathers in dissipative
and driven lattice systems do not occur in families of localised solutions as they are provided by discrete sets 
of attractors for appropriate initial conditions contained 
in the corresponding basin(s) of attraction \cite{Aubry1}-\cite{Maniadis}.

The aim of this work is to establish quantitative conditions in parameter space for 
the existence respectively non-existence of discrete breathers 
in general damped and driven anharmonic lattice systems. 
To this end we show that there exist parameter ranges such that for any  launched localised state 
 the difference 
between the maximal amplitude and the minimal amplitude of the oscillators decays exponentially fast. Consequently, 
a spatially uniform state is attained. Most importantly, this rules out the persistence of any non-uniform pattern 
such as breathers. 
Moreover, our results also identify parameter ranges for which
no inherent instability that is able to trigger the formation of a 
 localised pattern exists. Crucially, our rigorous  quantitative results 
establish prerequisites for the existence of discrete breathers in general damped and driven nonlinear lattices 
beyond the validity of the continuation process
starting from the anti-continuum limit  \cite{Marin},\cite{Aubry1}-\cite{Martinez2}. 
Our results are {\it generic} as they hold not only for any coupling strength but also 
for any on-site potential, any type of attractive interaction,
any degree of attractive interaction, general driving fields and arbitrary dimension of the system,.

We study the dynamics of general driven and damped nonlinear lattice systems of dimension $d$ given by the 
following system
\begin{eqnarray}
\ddot{q}_n&=&-U^{\prime}(q_n)-\gamma \dot{q}_n +A(t) + B(t) q_n\nonumber\\
&-& \sum_{j \in N_r(n)} \kappa_j \left[V^{\prime}(q_{n+j}-q_n)+V^{\prime}(q_{n}-q_{n-j})\right],
\label{eq:start}
\end{eqnarray}
with $n \in {\mathbb {Z}^d}$;\, $j,r\in {\mathbb {N}^d}$ and $N_r(n)$ is the set associated with the $r$ 
neighbours, $n+j$,  of site $n$ with $1\le j \le r$.
The variable $q_n(t)$ is 
the amplitude of the oscillator at site $n$ evolving in an anharmonic on-site potential 
$U(q_n)$.
The prime $^{\prime}$ stands for the derivative with respect to $q_n$ and an overdot $\dot{}$ 
represents the derivative with respect to time $t$.

The on-site potential $U$ is analytic  and is assumed to have the following properties:
\begin{equation}
U(0)=U^{\prime}(0)=0\,,\,\,\,U^{\prime \prime}(0)> 0.\label{eq:assumptions}
\end{equation}
In what follows we differentiate between soft on-site potentials and hard on-site potentials. For the former (latter)
the oscillation frequency of an oscillator moving in the on-site potential $U(q)$
decreases (increases) with increasing oscillation amplitude.
A soft potential possesses at least one inflection point. If a soft potential possesses a 
single inflection point, denoted by $q_i$, 
we suppose without loss of generality  (w.l.o.g.) that $q_i>0$. Then the following relations are valid
\begin{eqnarray}
U^{\prime}(-\infty <q<0)&<&0,\,\,\, U^{\prime}(0<q<q_i)>0,\\
U^{\prime \prime}(q_{i})&=&0,\,\,\,  U^{\prime \prime }(-\infty<q<q_i)>0.
\end{eqnarray} 
If $U(q)$ possesses two inflection points denoted by $q_{i,-}<0$ and $q_{i,+}>0$  it holds that 
\begin{eqnarray}
U^{\prime}(q_{i,-}<q<0)&<&0,\,\,\,  U^{\prime}(0<q<q_{i,+})>0\\
U^{\prime \prime}(q_{i,\pm})&=&0,\,\,\,  U^{\prime \prime}(q_{i,-}<q<q_{i,+})>0.\label{eq:U2prime}
\end{eqnarray}
We remark that $U(q)$ can have more than two inflection points (an example is a periodic potential $U(q)=-\cos(q)$). 
However, in the frame of the current study we are only interested in motion 
between the inflection points adjacent to the minimum of $U(q)$ at $q=0$.  
Hence, in the forthcoming we suppose that for soft on-site potentials the motion at each lattice site $n$
stays inbetween the inflection points, viz. $q_{i,-}<q_n(t)<q_{i,+}$, where $U(q)$ is convex.

Hard on-site potentials are, in addition to the assumptions in (\ref{eq:assumptions}), 
characterised in their entire range of definition  by 
\begin{equation}
U^{\prime}(q<0)<0,\,\,\,U^{\prime}(q>0)>0,\,\,\,U^{\prime \prime}(q)> 0.\label{eq:assumptionhard} 
\end{equation}
In contrast to soft potentials,  since the hard potentials are by assumption convex in their 
entire range of definition no boundedness condition as for the  motion in soft potentials is required.

Each oscillator interacts within the interaction radius $r$ with its neighbouring oscillators with (local) 
coupling strength $\kappa_j> 0$
(the interaction radius can range from  next neighbour coupling to global coupling) 
via forces derived from an attractive 
interaction potential $V(u)$ which 
is analytic and furthermore, is assumed to have the following features:
\begin{equation}
V(0)=V^{\prime}(0)=0,\,\,\, V^{\prime \prime}(0)\ge 0,\,\,\,  V^{\prime \prime}(u \ne 0)> 0.\label{eq:assumptions1}
\end{equation}
Thus $V(u)$ is convex which is further characterised  by $V^{\prime}(u>0)>0$ and $V^{\prime}(u<0)<0$. It is through the 
site-dependent coupling strength that heterogeneity enters the model.  The interaction potential 
can be harmonic but also anharmonic such as  for example in $\beta-$Fermi-Pasta-Ulam systems and Toda-type interactions. 

The parameter $\gamma >0$ regulates the strength of the damping. $A(t)$  
and $B(t)$ are smooth functions representing  general  external 
time-dependent fields with 
\begin{eqnarray}
 \max_{t \in \mathbb{R}} A(t) &=& A_{max}< \infty\,,\,\,\min_{t \in \mathbb{R}} A(t) =A_{min}>-\infty\nonumber,\\
  \max_{t \in \mathbb{R}} B(t) &=& B_{max}< \infty\,,\,\, \,\, \min_{t \in \mathbb{R}} B(t) =B_{min}> -\infty\nonumber.
\end{eqnarray}
The $A(t)$ and $B(t)$ term in Eq.\,(\ref{eq:start}) is associated with direct and parametric driving respectively.

We investigate under which circumstances  Eq.\,(\ref{eq:start})  possesses  
 time-periodic and spatially  localised solutions, viz. discrete breathers, $q_n(t+T_b)=q_n(t)$,  
with period $T_b=2\pi/\omega_b$ where $\omega_b$ denotes the breather frequency. 
We consider {\it all possible} standard breather solutions involving
single-site breathers as well as multi-site breathers in the following referred to as single breathers and multibreathers. 
While for 
the former all the oscillators perform inevitably in-phase motion for the 
latter the oscillators 
perform in-phase and/or out-of-phase periodic motion with respect to a 
reference oscillator \cite{Aubry1},\cite{Morgante}. Multibreathers can also consist of arrays of single breathers, 
viz. the pattern is localised around more than a single site or a single group of sites.   
Note that as one-dimensional lattices are concerned, it is proven in \cite{Cuevas},\cite{Koukouloyannis} that the only available stable multibreather solution 
are those with relative 
phase $0$ (in-phase) and $\pi$ (out-of-phase) between the  lattice sites and phase-shift breathers do not exist.  
Hence our treatment of breathers is comprehensive.

In general, breathers, being supported by periodic closed orbits in phase space,
are associated with periodic bounded motion of the oscillators inside their on-site potentials $U(q)$. 
Periodic solutions 
require time-periodic external fields $A(t)=A(t+T_A)$ and $B(t+T_B)$ with appropriate periods $T_A$ and $T_B$.

We introduce the following quantities related to the extremal values of the coordinates:
\begin{equation}
 q_{max}(t)=\max_n q_n(t),\qquad q_{min}(t)=\min_n q_n(t),\nonumber\\
\end{equation}
and denote the difference between them by 
\begin{equation}
 \Delta q(t)=q_{max}(t)-q_{min}(t)\ge 0\,.\nonumber
\end{equation}
The difference between the associated velocities is denoted by $\Delta \dot{q}(t)=\dot{q}_{max}(t)-\dot{q}_{min}(t)$.
In general for breather solutions with period $T_b$ it holds that $\Delta q(t)=\Delta q(t+T_b)$ and 
$\Delta \dot{q}(t)=\Delta \dot{q}(t+T_b)$.
Non-uniform (uniform) states are characterised by non-vanishing (identically vanishing) $\Delta q(t)$.

\noindent In the following we  list the {\bf conditions}  satisfied by breather solutions described above:

We first discuss single breather solutions being peaked around one lattice site and 
the oscillators perform in-phase motion. Later we comment on multibreathers. 

\noindent  The difference  $\Delta q(t)$ involves inevitably 
the same two oscillators  all the time. In fact, since for single breathers the pattern is spatially exponentially localised 
the two lattice sites involving 
$q_{max}(t)$ and $q_{min}(t)$ remain the same and only exchange their role 
after every change of sign of the periodically oscillating amplitudes. 
To be precise, the lattice sites $n=max$ and $n=min$ supporting the 
oscillators with $q_{max}$ and $q_{min}$ respectively during  phases when
$q_{n}\ge 0$ swap when the coordinates $q_n$ become negative. To 
describe the behaviour of $\Delta q(t)$ and $\Delta \dot{q}(t)$  we 
express a period duration $T_b=T_d+(T_b-T_d)$ as the sum of two stages of length $T_d$ and $T_b-T_d$ during which  
the coordinates $q_n(t)$ possess opposite sign. We first consider stages of length $T_d$ determined by
$k T_b\le t \le kT_b+T_d$ with $k=0,1,...$  
during which the coordinates are either non-negative or non-positive depending on the initial conditions. 
(We recall that $\Delta q$ is non-negative by definition.)  At the beginning of each interval 
the values are w.l.o.g. given by 
$\Delta q(kT_b)=\Delta q_0=0$ and $\Delta \dot{q}(kT_b)=\Delta \dot{q}_0>0$. (We remark that in the following the 
temporal evolution is considered 
 on such subintervals where $\Delta q(t)$ is smooth.)
Positive (negative) initial velocities $\dot{q}_n(kT_b)>0$ ($\dot{q}_n(kT_b)<0$) with 
$\dot{q}_{max}(kT_b)>\dot{q}_{min}(kT_b)>0$ ($\dot{q}_{min}(kT_b)<\dot{q}_{max}(kT_b)<0$)
result in non-negative (non-positive) amplitudes $q_n(t)\ge 0$ ($q_n(t) \le 0$) 
during intervals $kT_b \le t \le kT_b+T_d$.
That is, all oscillators are  at  $t=kT_b$  situated at the position
$q_n=0$, corresponding to the minimum position of the on-site potential, and 
$\Delta \dot{q}(kT_b)$ and $\Delta q(kT_b)$ 
attains its maximum and minimum respectively. During $kT_b \le t \le kT_b+T_d/2$, 
the quantity 
$\Delta \dot{q}(t)$  monotonically decreases resulting at $t=kT_b+T_d/2$ in 
$\Delta \dot{q}(t)=0$ while the monotonically increasing quantity $\Delta q(t)$ reaches its maximum. During 
$kT_b+T_b/2<t\le kT_b+T_d$ both $\Delta {q}(t)$  and $\Delta \dot{q}(t)<0$ monotonically decrease 
attaining at the end of the interval  $kT_b+T_d$ their 
minima $\Delta q(kT_b+T_d)=0$ and  $\Delta \dot{q}(kT_b+T_d)=-\Delta \dot{q}(kT_b)$. 

For the subsequent stage of length $T_b-T_d$, when the amplitudes 
$q_n(t)$ have opposite sign compared to the previous interval,  the motion of $\Delta q(t)$ and $\Delta \dot{q}(t)$  
starts with the same values as at the beginning of the previous interval, viz. 
$\Delta q_0=0$ and $\Delta \dot{q}_0>0$ and the oscillator at the lattice site 
that previously supported $q_{max}$ (and $\dot{q}_{max}$) 
possesses now the minimal amplitude $q_{min}$ (and minimal velocity $\dot{q}_{min}$) 
and vice versa.
However, $\Delta q(t)$ and $\Delta \dot{q}(t)$ 
resemble the  behaviour of their counterparts during the previous interval. 

As multibreathers are concerned the quantities $\Delta q$ and $\Delta \dot{q}$ 
exhibit qualitatively the same behaviour as for single breathers except 
that for phase differences $\pi$ the oscillators with $q_{max}> 0$ and $q_{min} <0$ possess opposite sign. 

In order to establish conditions for the non-existence of breather solutions we 
consider the behaviour of $\Delta q(t)$ and $\Delta \dot{q}(t)$ w.l.o.g.
on intervals  
\begin{equation}
I_{k}: k T_b+a\le t \le kT_b+T_d-a,\,\,\,{\rm with}\,\,\,k=0,1,2,...,\,\,\,{\rm and}\,\,\,0<a<\frac{T_d}{2}\label{eq:interval1}
\end{equation}
during which the 
amplitudes $q_n(t)$  are for single breathers and multibreathers 
with phase difference $0$ 
either exclusively non-negative or non-positive (see above) implying that the lattice site with $q_{max}$ is fixed and 
so is the lattice site with $q_{min}$.  For  multibreathers with phase difference $\pi$ between
the oscillators with $q_{max}$ and $q_{min}$ the same holds true  regarding the fixed positions of the extremal coordinates 
except that $q_{max}$ is always positive while $q_{min}$ is always negative. 
(For multibreathers more than one lattice site may support $q_{max}$ and/or $q_{min}$.) Note that $\Delta q(kT_b+a)=\Delta q(kT_b+T_d-a)>0$. For the forthcoming study it is appropriate to shift the time as $\tilde{t}=t-a$ shifting the intervals $I_{k}$ in (\ref{eq:interval1})
to
\begin{equation}
\tilde{I}_{k}: k T_b\le \tilde{t} \le kT_b+T_d-2a,\,\,\,{\rm with}\,\,\,k=0,1,2,...,\,\,\,{\rm and}\,\,\,0<a<\frac{T_d}{2}\label{eq:interval2}
\end{equation}
In what follows the tildes are omitted and  
at  $t=0$ the starting values $\Delta q_0$ and $\Delta \dot{q}_0$ are given by $\Delta q(0)=\Delta q_0>0$ and
$\Delta \dot{q}(0)=\Delta \dot{q}_0> 0$. 

$\Delta q(t)$ is smooth on the intervals $I_k$.  
Furthermore, on  each interval  $I_k$ it holds that
% \begin{eqnarray}
% \Delta q(t-(kT_b+T_d/2))&=& \Delta q(-t-(kT_b+T_d/2))\\ 
% \Delta \dot{q}(t-(kT_b+T_d/2))&=& -\Delta \dot{q}(-t-(kT_b+T_d/2))
% \end{eqnarray}
% with $k=0,1,...$, viz.
$\Delta q(t)$ is even with respect to $t_k=kT_b+T_d/2-a$ whereas $\Delta \dot{q}(t)$ 
is odd.  

Exploiting the symmetry features and periodicity of 
$\Delta q(t)$ and $\Delta \dot{q}(t)$ one obtains 
the following relations:
\begin{eqnarray}
\Delta q ((k+1)T_b)&=&\Delta {q}(kT_b+T_d-2a)=\Delta q(kT_b),\label{eq:Pmap1}\\
\Delta \dot{q}((k+1)T_b)&=&-\Delta\dot{q}(kT_b+T_d-2a)=\Delta\dot{q}(kT_b).\label{eq:Pmap2}
\end{eqnarray}
Crucially, the relations (\ref{eq:Pmap1}) and (\ref{eq:Pmap2}) constitute 
{\it necessary conditions} to be satisfied by  breather solutions. Thus, for given values $\Delta q(kT_b)$, 
$\Delta \dot{q}(kT_b)$ at the 
beginning of intervals $I_k$ 
 the solution $\Delta q(kT_b+T_d-2a)$, $\Delta \dot{q}(kT_b+T_d-2a)$ 
at the end of intervals $I_k$ can be utilised to derive a first 
recurrence (Poincar\'{e}) map $(\Delta q(jT_b),\Delta \dot{q}(jT_b)) \mapsto (\Delta q((j+1)T_b),\Delta \dot{q}((j+1)T_b))$ for which breathers constitute fixed points.  

The time evolution of the difference variable  $\Delta q (t)$ is determined by the following equation
\begin{eqnarray}
 \frac{d^2 \Delta q}{dt^2}&=& -\left[U^{\prime}(q_{max})-U^{\prime}(q_{min})\right]\nonumber\\
&-&\gamma (\dot{q}_{max}-\dot{q}_{min})+B(t)(q_{max}-q_{min})\nonumber\\
&-& \sum_{j \in N_r(n)}  \left\{\kappa_{max} \left[ V^{\prime}(q_{max+j}-q_{max})\right.
 +V^{\prime}(q_{max}-q_{max-j})\right]\nonumber\\
 &-&\kappa_{min}\left[ V^{\prime}(q_{min+j}-q_{min}) 
 +\left.V^{\prime}(q_{min}-q_{min-j})\right]\right\}\label{eq:deltap},
\end{eqnarray}
with 
\begin{equation}
 \kappa_{max}=\max_{1\leq j\leq r}\kappa_j\,;\,\,\,\kappa_{min}=\min_{1 \leq j \leq r}\kappa_j.
\end{equation}

Notice that the direct driving field $A(t)$ has no impact on $\Delta q(t)$. Regarding the maintenance of 
localisation the 
inequality $\Delta q (t)\ge  0$ constitutes  a necessary condition. Regarding the equal sign, 
for localised solutions, such as breathers, where the oscillators perform in-phase motion 
(and/or out-of-phase motion) $\Delta q(t)$ 
is zero only at instants of time   
when the oscillators, whilst performing periodic motion inside their potential wells, 
attain simultaneously the position $q_n=0$ at the minimum of the on-site potential. Conversely, 
if $\Delta q(t)$  decays approaching zero no localised pattern persists at all.

For the forthcoming derivations of estimates we facilitate the following statement:

\vspace*{0.5cm}

\noindent {\bf Lemma:}  For  soft potentials $U(q)$ with two
inflection points $q_{i,\pm}$ consider the interval 
\begin{equation}
I_s:=[q_l,q_r]\,\,\,{\rm with}\,\,\, q_{i,-}<q_{l},\,\, {\rm and},\,\,\, q_r<q_{i,+}\,.\label{eq:Is0}
\end{equation}.   

Then it holds that  for any pair $x,y \in I_s$ with $x< y$ 
\begin{equation}
\left[U^{\prime}(y)-U^{\prime}(x)\right] > \delta_s \left(y-x\right)>0\label{eq:deltabounds}
\end{equation}
where the constant $\delta_s>0$ is given by 
\begin{equation}
 \delta_s=\min_{q \in I_s}  U^{\prime \prime}(q)= \min\left[U^{\prime \prime}(q_{l}),U^{\prime \prime}(q_{r})\right].\label{eq:deltas}
\end{equation} 

For hard potentials consider the interval 
$I_h:=[x_l,x_r]$ with $-\infty<x_l$, $x_r<\infty$.
Then it holds that  for any pair $x,y \in I_h$ with $x< y$
\begin{equation}
\left[U^{\prime}(y)-U^{\prime}(x) \right] >\delta_h (y-x)>0\label{eq:deltaboundh}
\end{equation}
and the constant $\delta_h >0$  is given by 
\begin{equation}
 \delta_h=\min_{q \in I_h} U^{\prime \prime}(q)=U^{\prime \prime}(0).\label{eq:deltah}
\end{equation} 

\vspace*{0.5cm}

\noindent {\bf Proof:} Consider the expression
\begin{equation}
F(x,y)=\frac{ U^{\prime}(y)-U^{\prime}(x) }{y-x}.
\end{equation}
By assumptions (\ref{eq:U2prime}) and (\ref{eq:assumptionhard}) we have that on intervals $I_s$ and $I_h$ it holds that  
$U^{\prime}(y)>U^{\prime}(x)$ for $y>x$. Therefore the expression $F(x,y)$ is positive.
Furthermore, by virtue of the mean value theorem there exist a point $z$ in $(x,y)$ such that 
\begin{equation}
\frac{U^{\prime}(y)-U^{\prime}(x) }{y-x}=U^{\prime \prime}(z) \ge \min_{q \in I_s,I_h} U^{\prime \prime}(q).
\end{equation}
One has for  soft potentials  $\min_{q \in I_s}( U^{\prime \prime}(q))=
\min\left[U^{\prime \prime}(q_{l}),U^{\prime \prime}(q_{r})\right]>0$, and therefore 
it holds that 
\begin{equation}
 U^{\prime}(y)-U^{\prime}(x)\ge \min\left[U^{\prime \prime}(q_{l}),U^{\prime \prime}(q_{r})\right](y-x)=\delta_s (y-x)>0.
\end{equation}

Similarly for  hard potentials by the assumption (\ref{eq:assumptions}) one has 
$\min_{q \in I_h} (U^{\prime \prime}(q))=U^{\prime \prime}(0)>0$, 
so that \begin{equation}
 U^{\prime}(y)-U^{\prime}(x)\ge U^{\prime \prime}(0)(y-x)=\delta_h (y-x)>0.\nonumber
\end{equation}
completing the proof.

\hspace{16.5cm} $\square$

\vspace*{0.5cm}
Remark: To apply Lemma  to the case of  
soft potentials with a single inflection point $q_i>0$ one proceeds along the lines given above 
for the Lemma considering the interval $(-\infty,q_r]$ and $q_r<q_i$. The positive constant $\delta_s$ is given by
$\delta_s=\min_{q\in I_s}\left[U^{\prime \prime}(q)\right]=U^{\prime \prime}(q_{r})$.

\vspace*{0.75cm}

In the following we present conditions for which
 $\Delta q(t)$, associated with a breather solution satisfying the conditions listed above, 
exponentially decays which rules out the existence of  breather solutions to Eq.\,(\ref{eq:start}).  

\vspace*{0.5cm}
\noindent {\bf Theorem:}  
Let the relation $(\gamma/2)^2 > \omega_0^2-B_{max}>0$  be valid 
with $\omega_0^2=\delta_s$ and $\omega_0^2=\delta_h$ for soft and hard on-site potentials given in Eq.\,(\ref{eq:deltas}) and 
(\ref{eq:deltah}) respectively. 
Then it holds that Eq.\,(\ref{eq:start}) does not possesses
breather solutions.

\vspace*{0.5cm}
\noindent {\bf Proof:} We prove the assertion by contradiction. That is we  
suppose that Eq.\,(\ref{eq:start}) exhibits breather solutions associated with periodic functions 
$\Delta q(t+T_b)=\Delta q(t)$ and and $\Delta \dot{q}(t)=\Delta \dot{q}(t+T_b)$ 
satisfying the  necessary conditions in (\ref{eq:Pmap1}) and (\ref{eq:Pmap2}). 
Using the conditions in  (\ref{eq:assumptions}) and (\ref{eq:assumptions1})
together with the Lemma enables us 
to bound the r.h.s. of Eq.\,(\ref{eq:deltap}) on each of the intervals $I_{k}$, $k=0,1,..$, 
defined in (\ref{eq:interval2}),
from above  as follows: 
\begin{eqnarray}
 \frac{d^2 \Delta q}{dt^2}&=&-\left[U^{\prime}(q_{max})-U^{\prime}(q_{min})\right]
-\gamma (\dot{q}_{max}-\dot{q}_{min})\nonumber\\
&+&B(t)\left( q_{max}-q_{min}\right)\nonumber\\
&-& \sum_{j \in N_r(n)} \left[\kappa_{max}\left(\underbrace{V^{\prime}(q_{max+j}-q_{max})}_{\ge 0}
+ \underbrace{V^{\prime}(q_{max}-q_{max-j})}_{\ge 0}\right)\right.\nonumber\\
 &-&\kappa_{min}\left(\underbrace{V^{\prime}(q_{min+j}-q_{min})}_{\le 0}
 +\left.\underbrace{V^{\prime}(q_{min}-q_{min-j})}_{\le 0}
 \,\right) \right]\nonumber\\ 
 &\le& -\omega_0^2 \left(q_{max}-q_{min}\right) -\gamma (\dot{q}_{max}-\dot{q}_{min}) \nonumber\\
 &+&B_{max}\left(q_{max}-q_{min}\right)=
 -(\omega_0^2 -B_{max}) \Delta q\nonumber\\
 &-&\gamma \frac{d\Delta q}{dt}.\nonumber
\end{eqnarray}

Therefore, by the comparison principle for differential equations, $\Delta q(t)$ and $\Delta \dot{q}(t)$  are bounded from above by the solution of 
\begin{equation}
\frac{d^2 \Delta q}{dt^2}=-(\omega_0^2 -B_{max})\Delta q-\gamma \frac{d\Delta q}{dt}.\label{eq:boundhard}
\end{equation}
The solution to Eq.\,(\ref{eq:boundhard}) with initial conditions $\Delta q_{0,k}=\Delta q(kT_b)>0$, 
$\Delta \dot{q}_{0,k}=\Delta \dot{q}(kT_b)>0$ is given for  
$(\gamma/2)^2 >\omega_0^2-B_{max}>0$ on each interval $I_k$ 
by 
\begin{eqnarray}
 \Delta q_{k}(t)&=& \exp\left(-\frac{\gamma}{2} t\right) \left[
 \frac{\Delta \dot{q}_{0,k}+\frac{\gamma}{2}\Delta q_{0,k}}{W}\sinh(W t)\right.\nonumber\\
 &+&\left.\Delta q_{0,k}\cosh(W t)\right], \label{eq:decay}
\end{eqnarray}
and
\begin{eqnarray}
 \Delta \dot{q}_{k}(t)&=&  \exp\left(-\frac{\gamma}{2} t\right) 
 \left[ \left[ W\left(1-\left(\frac{\gamma}{2W}\right)^2\right)
 \Delta q_{0,k}\right.\right.\nonumber\\
&-&\left.\left.\frac{\gamma}{2W}\Delta \dot{q}_{0,k}\right]\sinh(W t)+ 
 \Delta \dot{q}_{0,k}\cosh(W t)\right] \label{eq:decayp}
\end{eqnarray}
where the index $k$ refers to the interval $I_k$ and $W=\sqrt{(\gamma/2)^2-(\omega_0^2-B_{max})}$.
Due to the Eqs.\,(\ref{eq:Pmap1}),(\ref{eq:Pmap2}), fulfilled by  breather solutions, 
the following recursion relations are true
\begin{eqnarray}
\Delta q_{0,k+1}&=&\Delta {q}_k\left(kT_b+T_d-2a\right)\\
 \Delta \dot{q}_{0,k+1}&=&-\Delta\dot{q}_k\left(kT_b+T_d-2a\right)
 \end{eqnarray}
 with starting values $\Delta q_{0,k=0}>0$ and  $\Delta \dot{q}_{0,k=0} > 0$ (see above).
Using the latter recursions and the notation
$Q_j=\Delta q(j T_b)$ and $P_j=\Delta \dot{q}(j T_b)$ with $j=0,1,...$ 
we cast the solution  in form of a first recurrence (Poincar\'{e}) map 
 \[ \left( \begin{array}{c}
 Q_{j+1} \\ P_{j+1}
         \end{array} \right)=
    M 
         \left( \begin{array}{c} Q_{j} \\ P_{j}
         \end{array} \right)
         \]
where the matrix $M_j$ is given by 
\[M=E
\left( \begin{array}{ll}
                 \left(C+\frac{\gamma}{2W}S\right) & \frac{1}{W}S \\
                 W\left(\left(\frac{\gamma}{2W}\right)^2-1\right)S & \left(\frac{\gamma}{2W}S-C\right)
                \end{array}
         \right)
         \]
with entries
\begin{eqnarray}
 E&=&\exp\left(-\frac{\gamma}{2} (T_d-2a)\right)\\
 C&=&\cosh\left(W (T_d-2a)\right)\\
 S&=&\sinh\left(W (T_d-2a)\right).
\end{eqnarray}
For the determinant of $M$ one obtains
\begin{equation}
\det M =-E^2\left(C^2-S^2 \right)=-E^2.
\end{equation}
As $\left|\det M\right|<1$ the Poincar\'{e} map is contractive and for any initial 
condition $\Delta q(0)$, $\Delta \dot{q}(0)$ the quantities 
$\Delta q(jT_b)$ and $\Delta \dot{q}(jT_b)$ exponentially decay and fall eventually below their initial values  
$\Delta q(0)=\Delta q_0>0$ and $\Delta \dot{q}(0)={\Delta}\dot{q}_0>0$ so that $\Delta q(t)$ and $\Delta \dot{q}(t)$ converge uniformly to zero 
which is in contradiction to the condition of 
periodic behaviour of non-vanishing $\Delta {q}(t)=\Delta q(t+T_b)$ and $\Delta \dot{q}(t)=\Delta \dot{q}(t+T_b)$ 
and the proof is complete.
  
\hspace{16.5cm} $\square$

Conclusively,  
our theorem provides conditions that 
rule out the existence and/or formation of breather solutions.

\vspace*{0.5cm}
\noindent {\bf Corollary:} Breather solutions to Eq.\,(\ref{eq:start}) can only  exist for 
\begin{equation}
 B_{max}>\omega_0^2.
\end{equation}

Remarkably, the process of exponential decay takes place {\it regardless of the amplitude of the external field} $A(t)$.
Furthermore,  exponential decay happens for {\it any kind of
attractive interaction potential} $V(u)$. As far as  hard on-site potential $U(q)$ is concerned, 
{\it only its curvature at the bottom, 
$U^{\prime \prime}(0)$, plays a role} for the decay process and the larger is 
the curvature the faster is the exponential decay while increasing the amplitude of the parametric driving $B_{max}$ 
has the opposite effect. Note that in order that the theorem applies the latter has 
to fulfill the constraint $B_{max}<\omega_0^2$. Importantly, the result holds for {\it general driving fields}.  
Interestingly, in our upper bound 
the decay rate turns out to be  independent of the initial distribution of the amplitudes and velocities $\{q_n(0)\}$ and 
$\{\dot{q}_n(0)\}$. 
They influence the amplitude of the decay of $\Delta q(t)$ and $\Delta \dot{q}(t)$ though. 

We stress that the hypothesis  $(\gamma/2)^2 > \omega_0^2 -B_{max}>0$ can be satisfied for 
arbitrarily small values of the damping strength $\gamma$ as 
for given $\omega_0^2=U^{\prime \prime}(0)$ for hard on-site potentials 
($\omega_0^2=\min[U^{\prime \prime}(q_l),U^{\prime \prime}(q_r)]$ for soft on-site potentials) the amplitude of the parametric driving 
field $B_{max}$ can be tuned to control the infimum of $\gamma$ complying with the inequality. 
Hence, $\gamma$ can be sufficiently small compared to a characteristic frequency of the system (which is e.g. given 
by oscillations near the bottom of a potential well 
with frequency determined by $U^{\prime \prime}(0)$) 
so that the system's dynamics is kept away from the overdamped limit.

Finally we remark that it is certainly of interest to extend the present study to systems 
that are discrete not only in space but also in time utilising the methods outlined in \cite{Hassan}.

\begin{figure}
\includegraphics[scale=0.75]{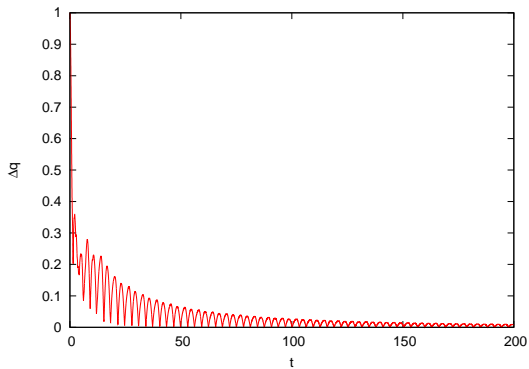}
\caption{Time evolution of $\Delta q(t)$ exhibiting on average exponential
decay in accordance with inequality 
(\ref{eq:decay}) for the system given in Eq.\,(\ref{eq:specific}).
The values of the parameters are $\gamma=0.35$, $A_0=B_0=0.99$, 
$\Omega_A=\Omega_B=2.0$, $\Theta_A=\Theta_B^0=0$, and $\kappa = 0.5$.} \label{fig:Deltaq}
\end{figure}
For an illustration of the exponential decay of an initially localised solution we choose for the hard on-site potential 
\begin{equation}
 U(q)=\frac{1}{2}q^2+\frac{1}{4}q^4.\nonumber
\end{equation}
The interaction potential is harmonic and is given by 
\begin{equation}
 V(q_n-q_{n-1})=\frac{1}{2}(q_n-q_{n-1})^2 ,\nonumber
\end{equation}
where the interaction radius is taken as $r=1$ amounting to linear
nearest-neighbour interaction and the coupling strength is uniform, i.e. 
$\kappa_n=\kappa$. 
As  the external fields are concerned we consider time-periodically varying fields and set 
for the direct driving  field $A(t)$ 
\begin{equation}
 A(t)=A_0 \sin(\Omega_A t+\Theta_A^{0})\nonumber
\end{equation}
with amplitude $A_0$, frequency $\Omega_A$ and phase $\Theta_A^0$. Similarly, for the parametric 
driving field $B(t)$  we choose
\begin{equation}
 B(t)=B_0 \sin(\Omega_B t+\Theta_B^0),\nonumber
\end{equation}
with amplitude $B_0$, frequency $\Omega_B$ and phase $\Theta_B^0$.

The corresponding lattice system is given by 
\begin{eqnarray}
\ddot{q}_n&=&-q_n-q_n^3+\kappa \left(q_{n+1}-2q_n+q_{n-1}\right)-\gamma \dot{q}_n\nonumber\\
&+&A_0\sin(\Omega_A t+\Theta_A^0)
+B_0 \sin(\Omega_B t+\Theta_B^0) q_n.
\label{eq:specific}
\end{eqnarray}
In our simulation the system comprises $N=100$ oscillators and periodic boundary conditions are imposed.
We plot in Fig.~\ref{fig:Deltaq} the temporal behaviour of $\Delta q(t)$  for the system (\ref{eq:specific}) 
starting from a localised single hump solution  peaked around the site $n=50$ 
associated with initial conditions $q_n(0)=1/\cosh(n-50))$ and $\dot{q}_n(0)=0$.
\begin{figure}
\includegraphics[scale=0.75]{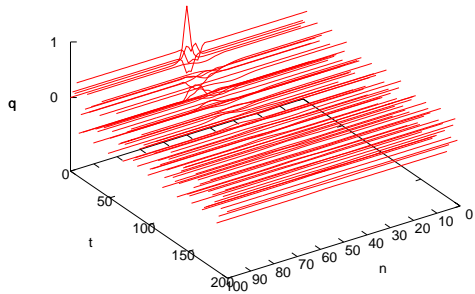}
\caption{Spatio-temporal pattern of the coordinates $q_n(t)$ for a chain consisting of $N=100$ oscillators with 
nearest-neighbour coupling. The parameter values are as given in Fig.~\ref{fig:Deltaq}.} \label{fig:spatio}
\end{figure}
$\Delta q(t)$  exponentially decays on average which is in accordance  with the inequality (\ref{eq:decay}) bounding the 
amplitude of $\Delta q(t)$ from above.
The corresponding spatio-temporal evolution of $q_n(t)$ is shown in Fig.~\ref{fig:spatio}
corroborating the  exponential decay of 
a spatial pattern. 
Eventually the dynamics settles on a spatially uniform state and the oscillators perform identical motion, 
$q_n(t)=q(t)$ and $\dot{q}_n(t)=\dot{q}(t)$ for all $n$, 
entailing that  the oscillators decouple. The oscillators undergo periodic motion on a limit cycle 
supporting periodic oscillations of the variables $q_n(t)$. 

In conclusion, we have studied the persistence and formation of non-homogeneous patterns represented 
by breather solutions 
in general nonlinear damped and driven
%one-dimensional  
lattice  systems.  Sufficient conditions, in terms of the values of the parameters, 
have been provided which assure that 
no time-periodic non-uniform state can exist. 
To be precise, it has been proven that  the difference between the maximal 
and minimal amplitudes of the lattice oscillators of a non-uniform time-periodic state decays exponentially 
fast. In this way we have proven that creation and/or preservation  of time-periodic, spatially (localised)  
patterns is impossible. Notably our results are independent of the number of oscillators and hold for arbitrary 
dimension of the system. 
Conversely,  rigorous {\it quantitative} conditions are identified under which 
discrete breathers can exist in general driven and damped lattices at all. Furthermore, our generic results on the 
non-existence of  time-periodic space-localised patterns and their formation in general nonlinear 
lattice systems open the way for further research on the emergent dynamical scenarios, in particular features of 
synchronisation, in coupled cell networks. Given that we have provided quantitative criteria in parameter space 
for the existence/nonexistence of discrete breathers the current work is also expected  to stimulate further 
experimental studies of breathers in nonlinear damped and driven lattice systems.

% \vspace{2.0cm}
% \noindent {\bf List of the figure captions}
% 
% \vspace{0.5cm}
% 
% \noindent Fig 1: Time evolution of $\Delta q(t)$ exhibiting on average exponential
% decay in accordance with inequality 
% (\ref{eq:decay}) for the system given in Eq.\,(\ref{eq:specific}).
% The values of the parameters are $\gamma=0.35$, $A_0=B_0=0.99$, 
% $\Omega_A=\Omega_B=2.0$, $\Theta_A=\Theta_B^0=0$, and $\kappa = 0.5$.
% 
% \vspace{0.25cm}
% \noindent Fig 2: Spatio-temporal pattern of the coordinates $q_n(t)$ for a chain consisting of $N=100$ oscillators with 
% nearest-neighbour coupling. The parameter values are as given in Fig.~\ref{fig:Deltaq}. 

\end{document}